\documentclass[12pt]{article}
\usepackage{amssymb}
\usepackage{epsfig}

\def\bbbone{{\mathchoice {\rm 1\mskip-4mu l} {\rm 1\mskip-4mu l}
{\rm 1\mskip-4.5mu l} {\rm 1\mskip-5mu l}}}

\newcommand{\beq}{\begin{equation}}
\newcommand{\eeq}{\end{equation}}
\newcommand{\beqn}{\begin{eqnarray}}
\newcommand{\eeqn}{\end{eqnarray}}
\newcommand{\bea}[1]{\beq\begin{array}{#1}}
\newcommand{\eea}{\end{array}\eeq}
\newcommand{\eq}[1]{(\ref{#1})}
\newcommand{\cT}{{\cal{T}}}
\newcommand{\dD}{{\cal D}}

\newcommand{\cH}{{\cal H}}

\newcommand{\intpi}{\int\limits^\pi_{-\pi}}
\newcommand{\intinf}{\int\limits^\infty_{-\infty}}

\newcommand{\dual}[1]{{}^{*}{#1}}

\newcommand{\tr}{\mathop{\rm Tr}}

\newcommand{\dd}{{\mathrm d}}
\newcommand{\Z}{Z\!\!\! Z}

\newcommand{\cC}{{\cal C}}
\newcommand{\cZ}{{\cal Z}}

\newcommand{\itep}
{~\vspace{-1.5cm}
\begin{flushright}
{\large ITEP-LAT/2003-16}\\
{\large KANAZAWA-03-22}
\end{flushright}
\vspace{1.0cm}}

\begin{document}
\baselineskip=14pt
\begin{center}

\itep

{\Large\bf Monopoles in Abelian Polyakov gauge and
projection (in)dependence of the dual superconductor mechanism of confinement}

\vskip 1.0cm
{\bf M.N.~Chernodub}
\vskip 4mm
{\it ITEP, B.~Cheremushkinskaya 25, Moscow, 117259, Russia and} \\
{\it Institute for Theoretical Physics, Kanazawa
University, Kanazawa 920-1192, Japan}

\end{center}

\begin{abstract}
We discuss the effect of the choice of an Abelian projection on
the dual superconductor mechanism of confinement in SU(2) gluodynamics.
Using qualitative arguments we show that the dual superconductor Lagrangian
corresponding to the Abelian Polyakov gauge has a different structure compared
to the dual Lagrangian in the Maximal Abelian gauge. A difference between
these Lagrangians reflects the fact that in continuum limit the monopoles should
be static in the Abelian Polyakov gauge and therefore these monopoles can not give rise
to the confinement of static quarks. Using the Abelian Polyakov gauge as an
example, we show that (i) the dual superconductor scenario of confinement
may depend on the Abelian projection;  (ii) the condensation of the Abelian
monopoles -- which may be realized in any Abelian projection -- is necessary
but not sufficient condition for confinement. These results do not exclude
an existence of a class of Abelian gauges in which the dual superconductor
scenario works well.
\end{abstract}

\vskip 0.3cm
{\bf 1.}
The confinement of color in QCD is one of the most interesting problems in quantum field theory.
Numerical lattice simulations~\cite{Bali:1994de,string:profile} of non--Abelian gauge theories clearly indicate
that the confinement of fundamental color charges is due to an appearance of
the chromoelectric string spanned between them. Despite an analytical explanation of
the string formation is still missing, the formation and properties of the chromoelectric string may be
studied within effective models. For example, the dual superconductor mechanism~\cite{DualSuperconductor}
provides a natural scenario of the string formation in QCD. This mechanism suggests that the vacuum
of QCD can be regarded as a media filled by condensed
Abelian monopoles. One of the properties of the monopole condensate is the dual Meissner effect
which squeezes the chromoelectric flux coming from the quarks into the flux tube. This flux tube
is analogous to the Abrikosov vortex in an ordinary superconductor.

The basic element of the dual superconductor mechanism is the Abelian monopole. This object
does not exist on the classical level in QCD. However, the monopoles can be identified with
particular configurations of the gluon fields by
the so--called Abelian projection formalism~\cite{AbelianProjections}.
This formalism relies on a partial gauge fixing of the SU(N) gauge symmetry up to an
Abelian subgroup. The Abelian monopoles appear naturally in the Abelian gauge as a result of
the compactness of the residual Abelian group.

Various numerical simulations indicate that the Abelian monopoles are likely to be responsible
for the confinement of quarks (for a review, see, $e.g.$, Ref.~\cite{Reviews}).
The Abelian monopoles provide a dominant contribution to the tension of the fundamental
chromoelectric string~\cite{AbelianDominance,ref:bali:bornyakov,shiba:string}.
The monopole condensate is formed in the low temperature (confinement)
phase and it disappears in the high temperature (deconfinement)
phase~\cite{MonopoleCondensation1,MonopoleCondensation2}.
The energy profile of the
chromoelectric string as well as the field distribution inside it can be described with a
good accuracy by the dual superconductor model~\cite{string:profile}.

Most of the results supporting the dual superconductor scenario were obtained in the
so called Maximal Abelian (MA) projection~\cite{kronfeld} of the SU(2) gauge theory.
This gauge is defined by the maximization of the lattice functional ($\sigma_i$ are the
Pauli matrices),
\beqn
\max_{\Omega} R_{MA}[U^{\Omega}]\,, \qquad
R_{MA}[U] = \sum_{s,\hat\mu}{\rm Tr}\Big(\sigma_3 U(s,\mu)
\sigma_3 U^{\dagger}(s,\mu)\Big)\,,
\label{R}
\eeqn
with respect to the gauge transformations,
$U(s,\mu) \to U^{\Omega}(s,\mu)=\Omega(s)U(s,\mu)\Omega^\dagger(s+\hat\mu)$.
In the continuum limit a local condition of the maximization can be written in the form
of the differential equation, $(\partial_{\mu}+igA_{\mu}^3)(A_{\mu}^1-iA_{\mu}^2)=0$.
Both the continuum gauge condition and the lattice functional~\eq{R} are invariant under
the residual U(1) gauge transformations,
\beqn
\Omega^{\mathrm{Abel}}(\omega) = {\mathrm{diag} (e^{i \omega},e^{- i \omega})}\,,
\label{eq:omega:abel}
\eeqn
where $\omega$ is an arbitrary function in continuum or on the lattice, respectively.

The MA gauge is a natural candidate for a realization of the dual superconductor
scenario because the MA gauge makes the off--diagonal gluon fields of freedom as
small as possible. Consequently, the diagonal (Abelian) components of the gluon
field are expected to play a leading role compared to the off-diagonal gluons and
the Abelian dominance~\cite{AbelianDominance} is a natural effect in the MA gauge.
The MA gauge belong to the class of extremization gauges which also includes the Minimal
Abelian gauge~\cite{ref:minopoles} defined by the minimization of the gauge functional~\eq{R},
and the Abelian gauges corresponding to minimization of the Abelian action and the Abelian monopole
density~\cite{Suzuki:1996ax}.

There is also a class of the diagonalization gauges which contains Abelian gauges
defined by the conditions of diagonalization of some functional $X[U]$ with respect
to gauge transformations $\Omega$. To define an Abelian gauge the functional $X[U]$
must belong to the adjoint representation of the $SU(2)$ gauge group~\cite{AbelianProjections}:
\beqn
   X[U] \to X[U^{(\Omega)}] = \Omega^\dagger X[U] \Omega\,.
\label{eq:diagonalization}
\eeqn
After the Abelian projection is fixed, the matrix $X[U]$ becomes diagonal
and the theory possesses the (residual) $U(1)$ gauge symmetry~\eq{eq:omega:abel}.
The  class of the diagonalization gauges includes, for example,
the Abelian Polyakov (AP), the Abelian field strength and the Abelian butterfly gauges. These gauges
correspond, respectively, to the diagonalization of the Polyakov line, the $U_{1,2}$ plaquette
variable and the butterfly operator~\cite{DiGiacomo:1999fa,DiGiacomo:1999fb}. The butterfly operator
is usually used in a simplest definition of the topological charge.

The success of the dual superconductor scenario in the MA gauge gives rise to the obvious
question~\cite{Suzuki:1994ay}: "Is the dual superconductor scenario gauge independent?".
The positive answer to this question
would imply that the monopoles identified in any Abelian gauge are always
associated with the confining configurations of the gluon field.
The dual superconductor mechanism would get an additional support in this case
because it is natural to think that the confinement -- as a gauge--invariant phenomenon
-- can not be described by the gauge--dependent model. On the other
hand the Abelian projection by itself can be considered as just a gauge--dependent tool to
associate the confining  gluon configurations with the Abelian monopoles. This tool may work well in
one gauge and may not work in another gauge.  Thus the negative answer to the above question
would not be a setback for the dual superconductor hypothesis either.
The negative answer would also imply that in one gauge the confining gluon
configurations may be associated with the monopoles while in another gauge these configurations
may be related to other objects. There are at lest two examples in favor of this way of thinking:
in the (indirect) Maximal Center gauge~\cite{ref:Greenisite}
the confining configurations are realized as center vortices, and in the Minimal Abelian
gauge~\cite{ref:minopoles} the confining configurations are probably associated
with the so-called "minopoles".

In the current literature there are conflicting opinions about the gauge independence of the
dual superconductor mechanism. The important indication of the gauge independence
is the observation of the Abelian and monopole dominance
not only in the MA gauge~\cite{AbelianDominance,ref:IR:monopoles}
but also in other Abelian gauges, for example in two Abelian projections corresponding to
minimization of the Abelian action and Abelian monopole density~\cite{Suzuki:1996ax}.
In the AP gauge the Abelian dominance holds automatically~\cite{Ejiri:1995vw}.
A generalization of the MA gauge, called the Laplacian
Abelian gauge, also possesses the Abelian and monopole dominance which are even stronger then in
the MA gauge~\cite{vanderSijs:1998jk}. Lattice studies indicate that in the MA gauge, in the AP
gauge and in the Abelian field strength gauge the vanishing of the Polyakov loop in the confinement
phase of SU(2) and SU(3) theories is due to the Dirac string associated with
the Abelian monopoles~\cite{Suzuki:1994ay}.

In Ref.~\cite{DiGiacomo:1999fa} the monopole condensation
in the SU(2) lattice gauge theory was numerically shown to be independent on the type of
the Abelian projection. Moreover in various Abelian projections the monopole condensate
vanishes at the temperature corresponding to the deconfinement phase transition.
Similar effect was also observed for the SU(3) gauge model~\cite{DiGiacomo:1999fb}.
The Abelian gauges used in Ref.~\cite{DiGiacomo:1999fa,DiGiacomo:1999fb} were the MA gauge,
the AP, Abelian field strength and Abelian butterfly gauges.
The gauge independence of the monopole condensate is also favored by the fact that
in the SU(2) gluodynamics the London penetration length measured in the MA projection is the same
as the one obtained without gauge fixing~\cite{Cea:1994ed}. Note also that recently proposed
gauge--invariant definition of the monopole~\cite{ref:fed} may also provide a gauge--independent description
of the dual superconductor.

On the other hand there are indications that the monopole dynamics is affected by the
choice of the Abelian projection. First of all, the length distribution of the monopole
trajectories is a gauge dependent quantity~\cite{Ejiri:1995vw}.
In the MA gauge a typical monopole ensemble consists of one large infrared
cluster and many short (ultraviolet) loops. The large monopole cluster (contrary to the
ultraviolet clusters) is responsible for the dominant contribution to the string tension coming
from the Abelian monopoles~\cite{ref:IR:monopoles}. In the extremization gauges such as the MA gauge
the infrared and ultraviolet clusters are clearly separable in histograms of the length
distribution while in the diagonalization gauges this is no longer the case~\cite{Suzuki:1996ax}.
Moreover, the density of the monopoles in the diagonalization gauges is typically higher
than in the extremization gauges. However, these results can not be considered as a
serious argument against the gauge independence since the above mentioned differences between
extremization and diagonalization gauges are due to the short monopole loops which are irrelevant to
the confinement of quarks~\cite{Suzuki:1996ax}. Moreover, the monopole density is not an order parameter
for the deconfining phase transition~\cite{DelDebbio:1991fp}.

An evidence against the gauge independence was given in Ref.~\cite{Bernstein:1996vr} where
the effect of the choice of the Abelian projection on the properties of the Abelian projected
chromoelectric string was studied. It was shown that the string in different projections
looks differently: the correlation length (the inverse monopole
mass) extracted from the string profile in the AP gauge is consistent with zero
contrary to the MA gauge~\cite{Bernstein:1996vr}.

Another indication of the projection dependence is based on measurements of the chiral condensate
in non--Abelian models with fermions.
It was shown in Refs.~\cite{Woloshyn:1994rv,Lee:1995ac} that the chiral condensate is dominated by
the contributions of the Abelian monopoles in the MA gauge of both SU(2) and SU(3) gauge theories.
However, this is no longer the case in the Abelian field strength~\cite{Woloshyn:1994rv} and the
AP~\cite{Lee:1995ac} gauges. It was checked in Ref.~\cite{Woloshyn:1994rv}
that in the field strength gauge the Abelian contribution to the chiral condensate does
not scale properly towards continuum and chiral limits because it
is insensitive to the lattice gauge coupling and to the quark mass.

Below we discuss the question of the projection (in)dependence of the dual superconductor scenario
considering an effective dual superconductor model in the AP gauge. We restrict ourselves
to the simplest case of the SU(2) gauge theory. We work in Euclidean space--time, and
we use both lattice and continuum formulations of this theory.

\vspace{5mm}
{\bf 2.} The main objects in the dual superconductor scenario are the Abelian monopoles
which are associated with singularities in the gauge fixing conditions. The easiest way
to understand the appearance of the Abelian monopoles is to consider the
diagonalization condition~\eq{eq:diagonalization}. If in some point $x$ of the space--time
the eigenvalues of the matrix $X$ coincide,
\beqn
X[x,A] = \left(
\begin{array}{cc}
\lambda_1(x) & 0\\
0 & \lambda_2(x)
\end{array}
\right)\,,
\qquad
\lambda_1(x) = \lambda_2(x)\,,
\label{eq:monopole:position}
\eeqn
then the gauge degrees of freedom can not be fixed
completely. The Abelian fields are singular in such a point. In order for the two
eigenvalues of the matrix $X[U]$ to coincide, three independent equations must be
satisfied~\cite{AbelianProjections}. Thus these
singularities form closed loops in the four-dimensional space--time.
It was argued in Ref.~\cite{AbelianProjections} that these loops correspond to trajectories of
the Abelian monopoles which should be considered as additional degrees of freedom in
the corresponding Abelian gauge. The closeness of the monopole loops indicates that the magnetic
charge is a conserved quantity.

Now let us study explicitly the AP projection. This projection can be considered
only in the case of non--zero temperatures when one of the directions of the
Euclidean space--time is compactified (we call it "time", or, "temperature", direction).
One can also study the zero--temperature case as a limit of arbitrarily small temperatures.

The AP projection is defined by the diagonalization condition~\eq{eq:diagonalization} where
the operator $X[U]$ coincides with the matrix $P[\vec s,s_4;U]$ trace of which is equal to
the Polyakov loop, $P[\vec s,U] = \frac{1}{2} \tr P[\vec s,s_4;U]$. On the lattice
this operator can be written as
\beqn
X[U] \equiv P[\vec s, s_4; U] = U_0(\vec s, s_4) \cdot U_0(\vec s, s_4+1) \dots U_0(\vec s, s_4-1)
\equiv \prod^{s_4-1}_{s'_4=s_4} U_0(\vec s, s'_4)\,.
\label{eq:PolyakovMatrix:lattice}
\eeqn
where the product of the lattice gauge fields $U_0$ goes along the closed parallel to the
time direction. The path starts and ends at the same point $(\vec x, x_4)$.

Let us show that in the continuum limit all the Abelian monopoles in the AP gauge must be static.
Here we follow Refs.~\cite{Chernodub:1997dr,Jahn:1998nw}. In the continuum limit the analog of the
matrix~\eq{eq:PolyakovMatrix:lattice} to be diagonalized is
\beqn
P_x[x,A] = \cT \exp \left\{ \frac{i}{2} \oint_{\cC_x} \dd x_0 \, \sigma_i \, A^i_0(x) \right\}\,,
\label{eq:PolyakovMatrix:continuum}
\eeqn
where the symbol $\cT$ means the path ordering and the integration goes over
the same closed path which starts and ends in the same point $x$.

Suppose, that a monopole trajectory passes through the point $x$. According to the
definition by 't~Hooft~\cite{AbelianProjections} in this point the
eigenvalues of the Polyakov loop coincide with each other. This immediately implies that
the matrix~\eq{eq:PolyakovMatrix:continuum} -- being an element of the SU(2) group --
must belong to the center of the $SU(2)$ group, $\Z_2$, $P[x,A] = \pm \bbbone$. Consequently,
the Polyakov loop is $P[\vec x] \equiv \frac{1}{2} \tr P[x,A] = \pm 1$. To prove the static
nature of the monopole current let us consider another point $y$ which lies on the same
Polyakov loop (in other words, $y_i = x_i$, $i=1,2,3$ while $y_4 \neq x_4$).
Due to the cyclic nature of the trace operation, $\frac{1}{2} \tr P[y,A]
\equiv \frac{1}{2} \tr P[x,A] = \pm 1$. Thus, the eigenvalues of the matrix
$P[y,A]$ in the point $y$ must also belong to the center of the group,
$P[y,A] = \pm \bbbone$. Thus, we conclude, that if an Abelian monopole
passes through the point $x=(\vec{x},x_4)$ it must also pass through all points $y$
with the same spatial coordinates:  $y=(\vec{x}, y_4)$ for all $y_4$. Thus
in the Polyakov Abelian projection all Abelian monopole trajectories
must be static\footnote{Note that a spatially degenerate configuration containing infinitely
closed points $\vec x_1$ and $\vec x_2$ such that $P[\vec{x}_1] = P[\vec{x}_2] \in \Z_2$
does not correspond to a monopole.}.

Similar derivation can be done on the lattice using the lattice definition of the
diagonalization matrix~\eq{eq:PolyakovMatrix:lattice} along with the 't~Hooft's condition of
the monopole positions~\eq{eq:monopole:position}. However, in real lattice simulations
the determination of the monopole positions using Eq.~\eq{eq:monopole:position} is problematic
because an exact equality of the Polyakov matrices~\eq{eq:PolyakovMatrix:lattice}
to an element from the center of the SU(2) group is a very rare event. This is expected
because the exact coincidence of a point--like monopole position with the lattice site
is highly improbable. Therefore a widely accepted way of the monopole localization on the lattice
uses a lattice version of the Gauss theorem known as the DeGrand--Toussaint (DGT)
construction~\cite{DGT}. In this construction the monopole positions are identified in
3D cubes which are sources of the magnetic flux. The disadvantage of this construction
is that in the AP gauge it produces many short monopole loops with the length of the order
of the lattice spacing $a$, as observed in Ref.~\cite{Suzuki:1996ax}. These loops couple
geometrically to each other and a large number
of the short loops makes a practical observation of the static monopole trajectories
impossible on the lattices with a moderate lattice spacing. However, one may naturally
expect that in the continuum limit, $a \to 0$, the DGT and 't~Hooft definitions of the
monopole loops should coincide with each other giving the static monopole trajectories only.

Note that the Polyakov loops in the AP gauge are Abelian by definition and the Abelian
dominance holds automatically in this gauge. However, the Abelian nature of these loops seems
to be inconsistent with the dual superconductor scenario because the key
objects in this scenario, the Abelian monopoles, are static in the AP gauge and therefore
these monopoles can not contribute to the Polyakov loop correlators.

\vspace{5mm}
{\bf 3.} Now let us derive an effective dual superconductor model in the AP gauge.
The standard dual superconductor model corresponding to the SU(2) gauge theory
in the $4D$ Euclidean space is described by the dual Ginzburg--Landau (DGL)
Lagrangian~\cite{ref:suzuki:maedan}:
\beqn
L_{DGL}[B,\Phi] = \frac{1}{4} F^2_{\mu\nu}
+ \frac{1}{2} {\bigl| D_\mu(B) \, \Phi\bigr|}^2 + V(\Phi)\,,
\label{eq:DGL}
\eeqn
where $F_{\mu\nu} = \partial_\mu B_\nu - \partial_\nu B_\mu$ is
the field strength of the dual gauge field $B_\mu$, $\Phi$ is the
monopole field with the magnetic charge $g_M$, and $D_\mu =
\partial_\mu + i g_M B_\mu $ is the covariant derivative. The
gauge field $B_\mu$ is dual to the third component of the gluon
field, $A^3_\mu$, in an Abelian gauge. The model possesses the dual $U(1)$ gauge
symmetry, $B_\mu \to B_\mu - \partial_\mu \alpha$, $\Phi \to e^{i
g_M \alpha}\, \Phi$. The form of the potential
\beqn
V(\Phi) = \lambda {\biggl({|\Phi|}^2 - \eta^2 \biggr)}^2\,,
\label{eq:V}
\eeqn
implies the existence of the monopole condensate, $|\langle\Phi\rangle| = \eta>0$.

To derive this effective model the authors of Ref.~\cite{ref:suzuki:maedan}
started from the partition function of the monopole currents:
\beqn
\cZ_{mon} = \int\hspace{-4.3mm}\Sigma \dD k \exp\Bigl\{
- \frac{g^2_M}{2} \int \dd^4 x \int \dd^4 y\, k_\mu(x) D(x-y) k_\mu(y)  - S_{int}(k)\Bigr\}\,,
\label{eq:Z:Coulomb}
\eeqn
where the first term in the action corresponds to the Coulomb interaction between the monopoles
($D(x)$ is the inverse Laplacian) and
$S_{int}(k)$ is the action of the closed monopole currents~$k$,
\beqn
k_{\mu}(x) = \oint \dd \tau\, \frac{\partial {\tilde x}_\mu(\tau)}{\partial \tau} \,
\delta^{(4)} [x -\tilde x(\tau)]\,.
\nonumber
\eeqn
Here the four--dimensional vector ${\tilde x_\mu}(\tau)$ represents the trajectory of
the monopole current~$k$.

Representing the Coulomb interaction in Eq.~\eq{eq:Z:Coulomb} as an integral over the
vector $B_\mu$, we get\footnote{Here and below
we omit the constant pre--factors in the partition functions.}
\beqn
\cZ_{mon} = \int\hspace{-4.3mm}\Sigma \dD k \int \dD B \, \exp\Bigl\{
- \int \dd^4 x\, \Bigl[\frac{1}{4} F^2_{\mu\nu} + i g_M \, k_\mu(x) \, B_\mu(x)\Bigr]
- S_{int}(k)\Bigr\}\,,
\label{eq:Zmon}
\eeqn
A further integration over monopole trajectories $k$ in Eq.~\eq{eq:Zmon} leads to the effective
Lagrangian~\eq{eq:DGL}. The potential term~\eq{eq:V} in Eq.~\eq{eq:DGL} comes from the
self--interaction of the monopole trajectories, $S_{int}(k)$, Ref.~\cite{ref:suzuki:maedan}.

The DGL model~\eq{eq:DGL} was shown to correctly describe the profiles of the chromoelectric
string~\cite{string:profile} as well as the monopole actions~\cite{Chernodub:2003pu}
obtained in numerical simulations of the lattice SU(2) gauge theory in the MA gauge.

The easiest way to derive the form of an effective action in the Polyakov gauge is to
use the formalism of differential forms in the lattice regularization described
in Ref.~\cite{diff:forms} and also in the second paper of Ref.~\cite{Reviews}.
Let us perform a lattice derivation of the
Lagrangian~\eq{eq:DGL} from Eq.~\eq{eq:Z:Coulomb}. Assuming for simplicity the quadratic form
of the self--interaction term, $S_{int}(k)$, we get the lattice version of Eq.~\eq{eq:Z:Coulomb}:
\beqn
Z = \sum_{\stackrel{\dual k \in \Z(\dual c_3)}{\delta \dual k = 0}} \exp\Bigl\{
- \frac{1}{2\beta} (\dual k, \Delta^{-1} \dual k) - \mu^2 \, {||\dual k||}^2 \Bigr\}\,,
\label{eq:Z1:latt}
\eeqn
where $\Delta^{-1}$ is the inverse lattice Laplacian and $\dual k$ is the
closed ($\delta \dual k =0$) monopole current defined on the dual lattice.
The scalar product of two forms $a$ and $b$ belonging to the cells $c$
($c$ = sites, links, plaquettes $etc.$) is denoted as $(a,b) = \sum_c a_c b_c$
and ${||a||}^2 \equiv (a,a)$. The parameter $\beta$ is the lattice gauge coupling
and the parameter $\mu$ defines the strength of the self--interactions of
the monopole currents.

Following Ref.~\cite{Polikarpov:1993cc} we can represent the partition
function~\eq{eq:Z1:latt} in the form:
\beqn
Z = \intinf \dD \dual B \intpi \dD \dual \varphi \intinf \dD \dual G
\sum_{\dual k \in \Z(\dual c_3)} \exp\Bigl\{ \hspace{-4mm}
& & - \frac{\beta}{2} {||\dd \dual B||}^2 - \frac{1}{4 \mu^2} \, {||\dual G||}^2 \nonumber\\
& & + i (\dual B + \dd \dual \varphi, \dual k) + i (\dual G, \dual k)\Bigr\}\,,
\label{eq:Z2:latt}
\eeqn
where we have introduced two Gaussian integrations, represented a closeness condition
as an integral over the compact scalar field $\varphi$, and used the lattice integration
by parts, $(\dd \dual \varphi, \dual k) \equiv (\dual \varphi, \delta \dual k)$.
Using the Poisson formula, $\sum_{m \in \Z} e^{i m x} \propto \sum_{n \in \Z} \delta(2 \pi m - x)$,
we get the constraint in Eq.~\eq{eq:Z2:latt}, $\dual G = \dd \dual \varphi + \dual B + 2 \pi \dual l$,
$\dual l \in \Z(\dual c_3)$. Integrating over the field $\dual G$ we get the following partition function,
\beqn
Z = \intinf \dD \dual B \intpi \dD \dual \varphi
\sum_{\dual l \in \Z(\dual c_3)} \exp\Bigl\{
- \frac{\beta}{2} {||\dd \dual B||}^2 - \frac{1}{4 \mu^2} \, {||\dd \dual \varphi + \dual B
+ 2 \pi \dual l||}^2\Bigl\}\,.
\label{eq:Z3:latt}
\eeqn
This is nothing but the lattice version of the dual Ginzburg--Landau model~\eq{eq:DGL} in the
London limit corresponding to the infinitely deep, $\lambda \to \infty$, potential~\eq{eq:V}
on the monopole field $\dual \Phi$. In this limit the radial part of the Higgs field is frozen and
only the phase $\dual \varphi$ of the monopole field is active. In Eq.~\eq{eq:Z3:latt} the interaction
of the phase $\dual \varphi$ of the monopole field with the dual gauge field is written in
the Villain form~\cite{Villain:1974ir}.

In the continuum limit the monopoles must be static in the AP gauge. To derive the AP analog
of the DGL model we impose the suppression of the spatial currents in the partition
function~\eq{eq:Z1:AP:latt}. To this end we introduce two parameters, $\mu_t$ and $\mu_{sp}$,
to control temporal and spatial currents, respectively:
\beqn
Z(\mu_t,\mu_{sp}) = \sum_{\stackrel{\dual k \in \Z(\dual c_3)}{\delta \dual k = 0}} \exp\Bigl\{
- \frac{1}{2\beta} (\dual k, \Delta^{-1} \dual k) - \mu^2_t \, {||\dual k||}^2_4
- \mu^2_{sp} \, {||\dual {\vec k} ||}^2 \Bigr\}\,,
\label{eq:Z1:AP:latt}
\eeqn
Here ${||\dual k||}^2_4 = \sum_{s} \dual k^2_{s,4}$ and
${||\dual {\vec k}||}^2 = \sum_{s} \sum_{i=1}^3 \dual k^2_{s,i}$ are the self--interactions of,
respectively, the temporal and the spatial currents. The AP projection corresponds to suppressed
spatial currents, $\mu_{sp} \to \infty$.

Applying to Eq.~\eq{eq:Z1:AP:latt} transformations which led us from Eq.~\eq{eq:Z1:latt}
to Eq.~\eq{eq:Z3:latt}) we get:
\beqn
Z = \intinf \dD \dual B \intpi \dD \dual \varphi
\sum_{\dual l \in \Z(\dual c_3)} \exp\Bigl\{
- \frac{\beta}{2} {||\dd \dual B||}^2 \hspace{-4mm}
& & - \frac{1}{4 \mu^2_t} \, {||\dd \dual \varphi + \dual B + 2 \pi \dual l||}^2_4 \nonumber\\
& & - \frac{1}{4 \mu^2_{sp}} \, {||\vec \dd \dual \varphi + \dual {\vec B} + 2 \pi \dual {\vec l}||}^2
\Bigl\}\,.
\label{eq:Z3:AP:latt}
\eeqn

One can immediately observe, that the suppression of the monopole currents corresponds
to the disappearance of the spatial part of the interactions between the monopole
field, $\dual \varphi$, and the dual gauge field, $\dual B$. Coming back to the continuum formulation
of the dual superconductor model we conclude that in the continuum AP gauge
the dual model should have the form
\beqn
L^{AP}_{DGL}[B,\Phi] & = & \frac{1}{4} F^2_{\mu\nu}
+ \frac{1}{2} {\bigl| D_4(B) \, \Phi\bigr|}^2 + V(\Phi) \nonumber\\
& \equiv & \frac{1}{4} F^2_{ij}
+ \frac{1}{2} F^2_{4i} + \frac{1}{2} {\bigl| D_4(B) \, \Phi\bigr|}^2 + V(\Phi)
\,.
\label{eq:DGL:AP}
\eeqn
The difference between this model and Eq.~\eq{eq:DGL} is in the spatial interaction
of the gauge and the monopole fields. Effectively, the spatial and temporal degrees of freedom
decouple from each other in the AP gauge. The temporal component of the gauge field, $B_4$, is
interacting with the monopole field while the spatial components, $B_i$, $i=1,2,3$, are free.
From the point of view of the original non--Abelian theory
the space--time asymmetry in the dual model~\eq{eq:DGL:AP} does not come as a surprise because
even in the limit of the infinitely large fourth dimension the AP gauge condition ($i.e.$, the
diagonalization of the matrix~\eq{eq:PolyakovMatrix:continuum}) violates the Lorentz symmetry.

\vspace{5mm}
{\bf 4.} Let us now discuss the consistency of the dual model~\eq{eq:DGL:AP} with known
numerical results in the AP gauge of the SU(2) lattice gauge theory. First, we evaluate the
monopole contribution to the quark potential.
The quantum average of the Abelian Wilson loop corresponds to the 't~Hooft loop in the dual
representation~\eq{eq:DGL:AP}:
\beqn
\cH_{\cC}[B] = \exp\Bigl\{ - \frac{1}{4} \int \dd^4 x\, \Bigl[ \Bigl(F_{\mu\nu} -
\frac{2\pi}{g_M} \dual \Sigma_{\mu\nu}\Bigr)^2  - F_{\mu\nu}^2\Bigr]\Bigr\}\,,
\label{eq:tHooft}
\eeqn
where $\dual \Sigma^\cC_{\mu\nu} = \frac{1}{2} \varepsilon_{\mu\nu\alpha\beta}
\Sigma^\cC_{\alpha\beta}$. The two--dimensional $\delta$--function,
\beqn
\Sigma^\cC_{\mu\nu}(x) = \int \dd^2 \tau \, \frac{\partial_{[\mu}
{\tilde x}_{,\nu]}}{\partial \tau_1 \partial \tau_2} \, \delta^{(4)} (x - {\tilde x}(\tau))\,,
\nonumber
\eeqn
represents a surface spanned on the trajectory of the particles $j_\mu^\cC$,
$\partial_\mu \Sigma^\cC_{\mu\nu} = j^\cC_\mu$. The surface is parameterized by the
vector ${\tilde x}(\tau_1,\tau_2)$. The quantum average of the
't~Hooft operator~\eq{eq:tHooft} does not depend on a particular the shape of the
surface $\Sigma^\cC$.

The 't~Hooft operator~\eq{eq:tHooft} effectively shifts the field strength,
$F_{\mu\nu} \to F_{\mu\nu} - \frac{2\pi}{g_M} \dual \Sigma_{\mu\nu}$. For static
quark and anti--quark located, respectively, at positions $(0,0,0)$ and $(R,0,0)$
of the three-dimensional space, the surface $\Sigma^\cC$ can be taken static and flat:
\beqn
\Sigma^\cC_{41}(x) = - \Sigma^\cC_{14}(x) = \Theta(x_1) \, \Theta(R-x_1) \,
\delta(x_2) \, \delta(x_3)\,.
\eeqn
All other components of the surface delta--function $\Sigma_{\mu,\nu}$ are equal to zero.
Therefore, the dual surface $\dual \Sigma^\cC$ has only two non--zero elements,
$\dual \Sigma^\cC_{23} = - \dual \Sigma^\cC_{32} = \Sigma^\cC_{01}$, which belong to the
three-dimensional part of the dual model~\eq{eq:DGL:AP}. The quantum average of such 't~Hooft
operator can easily be taken:
\beqn
\langle\cH_{\cC}\rangle  = {\mathrm{const.}}\, \exp\{- T V(R)\}\,,\quad
V(R) = \frac{g^2}{2}\, D_{3D}(R)\,,
\eeqn
where $D_{3D}(R)$ is the three dimensional inverse Laplacian,
$T$ is the length of the quarks trajectories, $T \gg R$, and
the prefactor does not depend on distance between the quarks, $R$. The electric charge of
the quark, $g$, is related to the magnetic charge of the monopole, $g_M$,
according to the Dirac condition, $g \cdot g_M = 2 \pi$.

Thus, in the dual model corresponding to the AP gauge the static quarks are not confined.
This is expected, because the monopoles are static in this gauge as we have already
discussed. In numerical simulation the string between the static color charges should be seen
as an infinitely thin object in the AP gauge because in the Lagrangian~\eq{eq:DGL:AP} the
spatial components of the dual gauge field $B_i$ do not couple to the monopole field, $\Phi$, and,
consequently, the monopole field does not feel the electric flux of the string.
This prediction agrees nicely with the observation of Ref.~\cite{Bernstein:1996vr}, where the
thickness of the Abelian string in terms of the monopole field ($i.e.$, the correlation length)
was found to be consistent with zero in the AP gauge.

Another interesting observation is that the fact of the absence of the confinement for the
static quarks in the AP gauge is correct for any form of the potential of the monopole
field, $V(\Phi)$. Indeed, the crucial fact of our derivation was the specific form of
the coupling of the monopole field to the dual gauge field in the Lagrangian~\eq{eq:DGL:AP}
while a particular form of the potential $V(\Phi)$ has not played any role.
Thus, the presence of the monopole condensate does not mean by itself that the quarks must be
confined. The interaction between the monopole field and the dual gauge fields are also essential
for the dual superconductor scenario of the confinement.

\vspace{5mm}
{\bf 5.} Summarizing, we have proposed a form for an effective Lagrangian of the dual
superconductor in the Abelian Polyakov gauge of the SU(2) gauge theory. Our derivation
has a qualitative nature based on the fact that the monopoles in this gauge must be static
in the continuum space--time. The monopole condensation -- as it follows from our study -- is
necessary but not sufficient condition for the explanation of the confinement phenomena within
the dual superconductor models.

Another consequence of our study is that the dual superconductor scenario of confinement
is projection--dependent\footnote{Note that our conclusion on the projection dependence is based
only on the comparison of the MA and AP gauges. Thus, our results do not exclude a possible existence of a class of
Abelian gauges (including the Maximal Abelian gauge) in which the dual superconductor scenario works well.}.
This statement does not contradict the observation of the independence of the monopole condensation on
the choice of the Abelian projection~\cite{MonopoleCondensation2,DiGiacomo:1999fb,DiGiacomo:1999fa},
because -- as we have shown -- the monopole
condensation by itself does not guarantee the confinement of color. On a technical level, the fact of existence
of the monopole condensate relies on the form of the potential of the monopole field, while
the quark--anti-quark potential is also dependent on other terms in the dual superconductor Lagrangian.
It seems that the most complete information about the monopole Lagrangian in a particular Abelian projection
can be extracted from the effective monopole action on the lattice~\cite{chernodub}. The monopole
action may further be used~\cite{Chernodub:2003pu,ref:blocking} to determine the form and the parameters of the dual
superconductor Lagrangian in the continuum limit.

These results do not exclude
the existence of a wide class of Abelian gauges (including the Maximal Abelian
gauge) in which the dual superconductor scenario works well.

Since the monopoles in the AP gauge are static they may contribute
to the string tension corresponding to the spatial Wilson loops. Moreover,
the Pontryagin index of the gauge field is related to the
magnetic charges defined in a axial version of the AP gauge~\cite{Reinhardt:1997rm}. Therefore,
in this gauge the magnetic monopoles are related to the nontrivial topological structures of
a non--Abelian gauge theory (a generalization of this statement to other Abelian
gauges can be found in Ref.~\cite{Jahn:1999wx}). Another interesting fact about the
Abelian monopoles in the AP gauge is that on a classical level they
are correlated~\cite{VanBaal:2001pm} with the positions of
the constituent monopoles in the van Baal--Kraan calorons~\cite{Kraan:1998pm}. Thus, the fact that
the monopoles in the Abelian Polyakov gauge are not relevant to the confinement of static quarks
does not mean that these monopoles do not carry interesting information about non--perturbative physics.

\section{Acknowledgments}
The author is grateful to A.~Di~Giacomo, F.~V.~Gubarev, T.~Suzuki and M.~I.~Polikarpov
for useful discussions. This work was supported by JSPS Fellowship No. P01023.

\end{document}